\documentclass[a4paper,11pt]{article}
\usepackage{pos}
\pdfoutput=1

\renewcommand{\vec}[1]{\boldsymbol{#1}}
\newcommand{\D}{\mathrm{d}}
\newcommand{\E}{\mathrm{e}}

\newcommand{\closer}{\hspace{-0.2ex}}
\newcommand{\NNj}{{\closer N\closer N_{\closer j}}}
\newcommand{\slopeB}{\beta}

\newcommand{\eg}{\emph{e.g.}}

\newcommand{\eVdist}{\kern-0.06em}
\title{Revisiting the AMS-02 antiproton excess: The role of correlated errors}
\ShortTitle{Revisiting the AMS-02 antiproton excess}

\author*[a,b]{Jan Heisig}
\author[c]{Michael Korsmeier}
\author[c]{Martin Wolfgang Winkler}

\affiliation[a]{Institute for Theoretical Particle Physics and Cosmology,
 RWTH Aachen University, Sommerfeldstr. 16, D-52056 Aachen, Germany}

\affiliation[b]{Centre for Cosmology, Particle Physics and Phenomenology (CP3), 
Universit\'e catholique de Louvain, Chemin du Cyclotron 2, B-1348 Louvain-la-Neuve, Belgium}

\affiliation[c]{The Oskar Klein Centre for Cosmoparticle Physics, Department of Physics, Stockholm University, Alba Nova, 10691 Stockholm, Sweden\vspace{3ex}}

\emailAdd{heisig@physik.rwth-aachen.de}
\emailAdd{michael.korsmeier@fysik.su.se}
\emailAdd{martin.winkler@su.se}

\abstract{Cosmic-ray antiprotons are a remarkable diagnostic tool for the study of astroparticle physics' processes in our Galaxy. While the bulk of measured antiprotons is consistent with a secondary origin, several studies have found evidence for a subdominant primary component in the AMS-02 data. In this proceedings article, we revisit the excess considering systematic errors that could affect the signal. Of particular importance are unknown correlations in the AMS-02 systematic errors, the dominant of which are associated with the cross sections for cosmic-ray absorption in the detector. We compute their correlations in a careful reevaluation of nuclear scattering data, utilizing the Glauber-Gribov theory to introduce a welcomed redundancy that we explore in a global fit. The inclusion of correlated errors has a dramatic effect on the significance of the signal. In particular, the analysis becomes more sensitive to the diffusion model at low rigidities. For a minimal extension beyond single-power-law diffusion, the global significance drops below $1\sigma$ severely questioning the robustness of the finding.
}

\FullConference{37$^{\rm{th}}$ International Cosmic Ray Conference (ICRC 2021)\\
		July 12th -- 23rd, 2021\\
		Online -- Berlin, Germany}

\begin{document}
\maketitle

%===================================================================
\section{Introduction}
%===================================================================

Cosmic-ray antiprotons are promising messengers for the study of astroparticle physics' processes in our Galaxy.
The bulk of measured antiprotons is consistent with a secondary origin, coming from scatterings of primary cosmic rays off the interstellar gas in the stellar disk. However, the precise data from the AMS-02 experiment~\cite{Aguilar:2016kjl} allows us to search for a subdominant primary component of antiprotons, \eg~from dark matter annihilation.

Several groups have reported a statistical preference for such a component in the rigidity range $10\!-\!20$\,GV which is compatible with a dark-matter signal~\cite{Cuoco:2016eej,Cui:2016ppb,Cuoco:2017rxb,Reinert:2017aga,Cui:2018klo,Cuoco:2019kuu,Cholis:2019ejx,Lin:2019ljc}. 
It hints at a dark-matter mass $50\:\text{GeV}\lesssim m_\text{DM}\lesssim100\:\text{GeV}$ and an annihilation cross section consistent with thermal freeze-out, $\langle \sigma v\rangle\!\sim \!10^{-26}\:\text{cm}^3/\text{s}$. Intriguingly, dark matter with almost identical properties has also been considered as an explanation of the gamma-ray Galactic center excess~\cite{Goodenough:2009gk}. 
However, while the various independent analyses agree on the preferred dark-matter properties the significance of the excess is highly controversial ranging from around $1\sigma$ to above $5\sigma$, see \cite{Heisig:2020jvs} for a recent review on the subject.

The discrepancy could point to unaccounted systematic uncertainties that affect the different analyses to a different extent or mimic the excess altogether.
One of the leading uncertainties comes from antiproton production cross sections that enter the background prediction. Their inclusion has, indeed, reduced the antiproton excess in~\cite{Reinert:2017aga} while its effect has been found to be marginal in~\cite{Cuoco:2019kuu}.

Another important source of correlated uncertainties comes from the AMS-02 measurement itself. In the rigidity range of interest, systematics dominate which could exhibit  correlations. While no covariance matrix has been provided by the collaboration, their inclusion can have a dramatic effect on the significance of the signal as shown in~\cite{Cuoco:2019kuu}. 
The most relevant source of correlated errors in the rigidity range of interest is the uncertainties in the cross sections for cosmic-ray absorption in the detector the reported fluxes are corrected for. Measurements of the involved inelastic nucleon-nucleus cross sections are sparse and often subject to considerable systematics. Furthermore, error correlations were unknown. 

Here we summarize recent results~\cite{Heisig:2020nse} addressing this lack of knowledge. We employ the Glauber-Gribov theory of inelastic scattering~\cite{Glauber1959,Gribov:1968jf} to provide a prediction for the absorption cross section based on the input quantities of the model, namely the nucleon-nucleon scattering cross sections and nuclear density functions. These quantities are subject to independent measurements and, hence, introduce a welcomed redundancy that allows us to reduce the uncertainties and compute the error correlations via a global fit of a large experimental dataset.
Including the systematic error correlation, we perform a spectral search for dark matter in the AMS-02 antiproton data.\footnote{For brevity, we concentrate on the cosmic-ray propagation setup corresponding to~\cite{Cuoco:2019kuu}. However, reference~\cite{Heisig:2020nse} also includes results for the complementary propagation setup used in~\cite{Reinert:2017aga}.}
This allows us to draw more robust conclusions on the existence of the antiproton excess and point to the decisive quantities its statistical significance depends on.

The remainder of this paper is organized as follows. In Sec.~\ref{sec:systematic} we summarize the various sources of systematic errors subject to correlations.
In Sec.~\ref{sec:absorption} we present our results for the nucleon-nucleus absorption cross sections and the corresponding correlation matrix. Finally, in Sec.~\ref{sec:excess} we discuss the implications for the tentative signal before drawing our conclusions in Sec.~\ref{sec:concl}.

%===================================================================
\section{Systematic uncertainties in the AMS-02 data}\label{sec:systematic} 
%===================================================================

\begin{figure*}[t]
\centering
\setlength{\unitlength}{1\textwidth}
\begin{picture}(0.99,0.42)
 \put(-0.0055,-0.02){\includegraphics[width=0.53\textwidth, trim= {3.3cm 2.2cm 3cm 0.8cm}, clip]{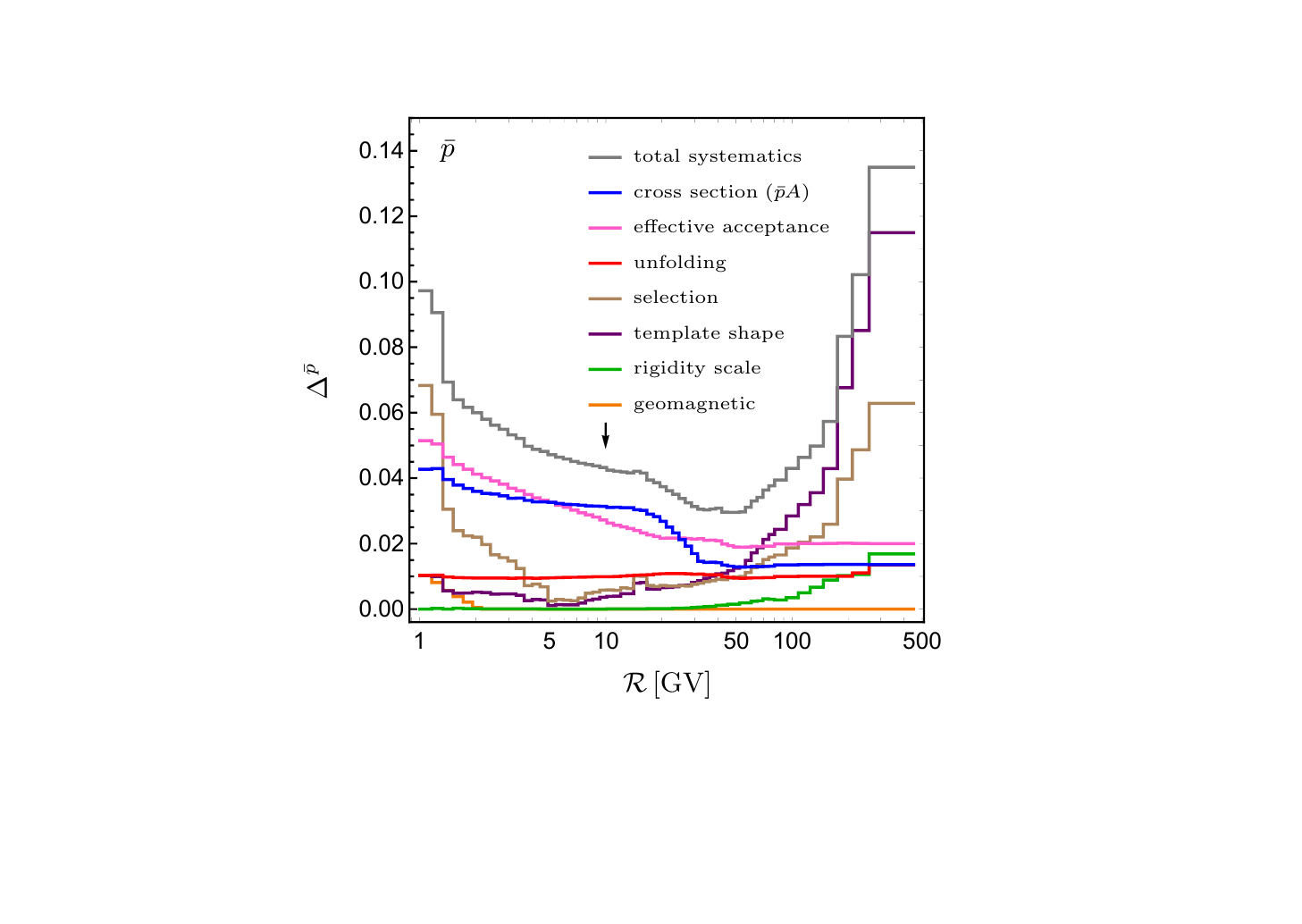}}
 \put(0.506,-0.02){\includegraphics[width=0.53\textwidth, trim= {3.3cm 2.2cm 3cm 0.8cm}, clip]{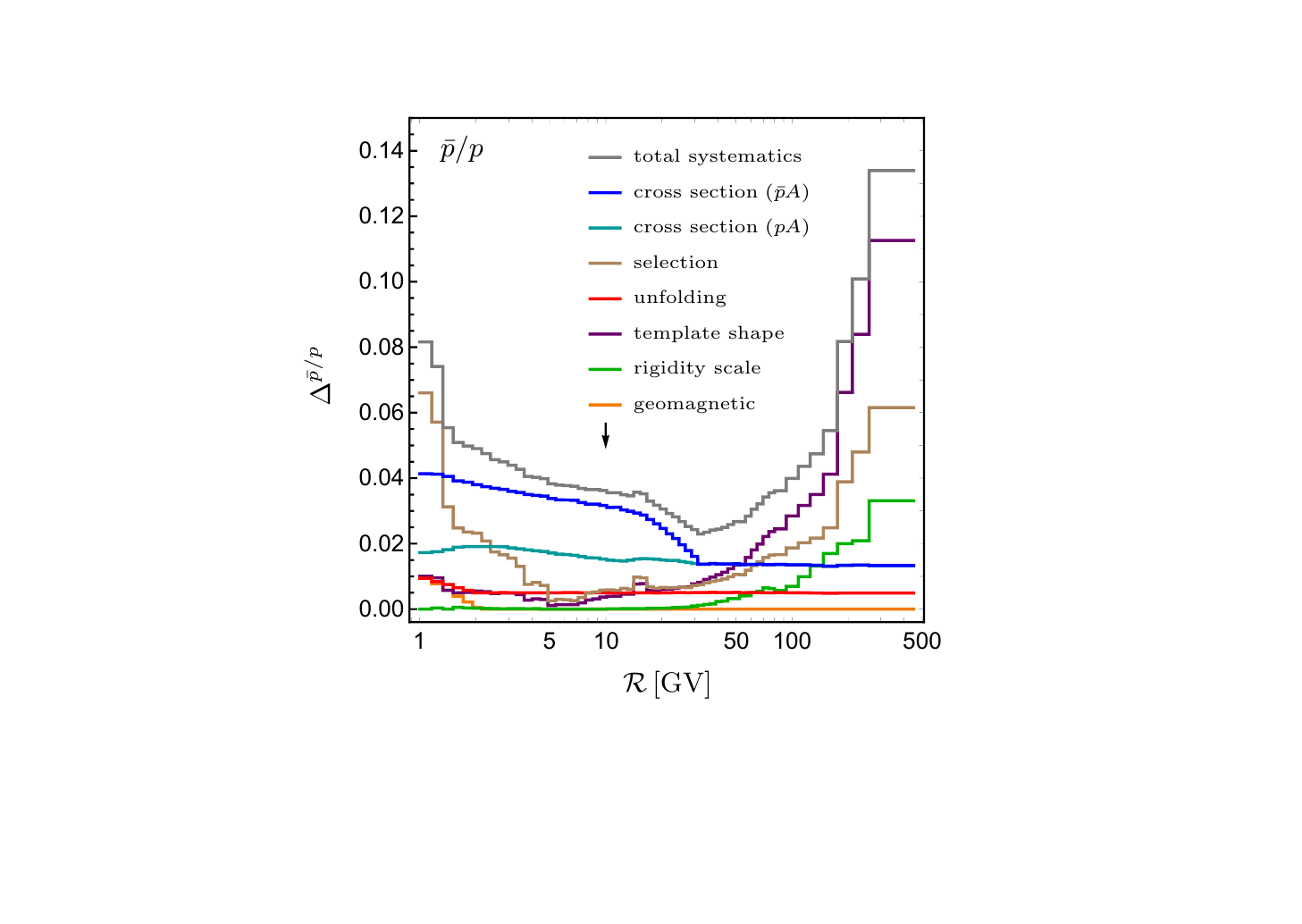}}
\end{picture}
\caption{
Relative systematic errors of the AMS-02 $\bar p$ (left) and $\bar{p}/p$ data (right). The contributions in the legend are ordered according to their size at ${R \cal}=10$\,GV as indicated by the arrow. Taken from~\cite{Heisig:2020nse}
\label{fig:systematicerrors}
}
\end{figure*}

In the rigidity range ${\cal R}\sim2\!-\!40$\,GV, systematics dominate the uncertainties in the AMS-02 antiproton data.
Figure~\ref{fig:systematicerrors} shows all sub-contributions of the systematic error in the $\bar{p}$ flux (left) and $\bar{p}/p$ flux ratio (right).
The most relevant contributions come from cross sections for cosmic-ray absorption in the detector. The reported fluxes are corrected by the corresponding loss through interactions with the detector material in the upper layers. The computation of the associated error correlations is one of the main subjects of the work presented here and is discussed in Sec.~\ref{sec:absorption}.

%=====================
%    \                                           |
%      \                                         |
%        \                                       |
%                                      \         |
%                                        \       |
%                                          \     |
%=====================

For the $\bar{p}$  flux, the effective acceptance error is of similar relevance. 
It is a residual systematic error in the effective folded acceptance and is estimated by a comparison of Monte Carlo simulation and direct measurements in several parts of the detector. As it affects antiprotons and protons in the same way, it cancels out in the $\bar{p}/p$ ratio. A data-driven analysis performed in~\cite{Heisig:2020nse} implies that the effective acceptance error is subject to short-range correlations (in rigidity).

Further systematics play a subleading role being small compared to either the previous systematics or the statistical error. They comprise the unfolding error, associated with the migration of events into neighboring rigidity bins, the template shape and selection error, arising from the choice of the template shape and cuts on the track shape, the rigidity scale error addressing misalignments of the tracker planes and uncertainties in the magnetic field map, and the error associated to the geomagnetic cut-off. For these contributions, we adopt the correlation lengths estimated in~\cite{Boudaud:2019efq}.

%===================================================================
\section{Global fit of nucleon-nucleus absorption cross sections}\label{sec:absorption}
%===================================================================

Within the Glauber-Gribov model of inelastic scattering~\cite{Glauber1959,Gribov:1968jf}  the nucleon-nucleus absorption cross sections can be expressed as
\begin{equation}
\label{eq:sigmaabs}
\sigma_\text{abs}^{N\closer A} (p_\text{lab})
=\int \!\D^2 b \left[1 - \left(1- \frac{2\,\text{Im}\chi_N(\vec b,p_\text{lab})}{A}\right)^{\!A\,} \right] + \Delta \sigma_\text{inel.\,screen.}(p_\text{lab})
\end{equation}
where $\vec b$ is the impact parameter, $N$ and $A$ denote the incident nucleon and the atomic mass number of the nucleus, respectively, and
\begin{equation}
\label{eq:chisph}
\text{Im} \chi_N(b,p_\text{lab}) = \frac12 \sum_{j=1}^A\sigma_\NNj \closer(p_\text{lab})\int_0^\infty\!\!\!\D q\, q\,J_0(b q) \,\E^{-\slopeB_\NNj \closer(p_\text{lab})\,q^2/2} 
\int_0^\infty \!\!\!\D r\, r^2  J_0(r q)\,\rho_j(r)\,,
\end{equation}
is the (imaginary part of the) phase-shift function assuming a spherical symmetric nuclear density function, $\rho(r)$. The input quantities of this model are the 
total nucleon-nucleon cross sections, $\sigma_\NNj\closer(p_\text{lab})$, the slope of the respective differential inelastic cross sections in the forward direction, $\slopeB_\NNj\closer(p_\text{lab})$, and the parameters of the nuclear density function chosen to be the harmonic oscillator shell model. The sum in eq.~\eqref{eq:chisph} runs over all constituents, $N_j$, of the nucleus and $J_n$ denotes the Bessel function of the first kind. The term $\Delta \sigma_\text{inel.\,screen.}(p_\text{lab})$ accounts for inelastic screening corrections~\cite{Karmanov:1973va}, see~\cite{Heisig:2020nse} for more details.
We use analytical parametrizations for $\sigma_\NNj\closer(p_\text{lab})$, $\slopeB_\NNj\closer(p_\text{lab})$~\cite{Heisig:2020nse} fitted to the cross-section data collected in Ref.~\cite{Tanabashi:2018oca,Okorokov:2015bha}.

\begin{figure*}[t]
\centering
\setlength{\unitlength}{1\textwidth}
\begin{picture}(0.5,0.32)
 \put(-0.0055,-0.03){\includegraphics[width=0.53\textwidth, trim= {3.3cm 2.2cm 3cm 2cm}, clip]{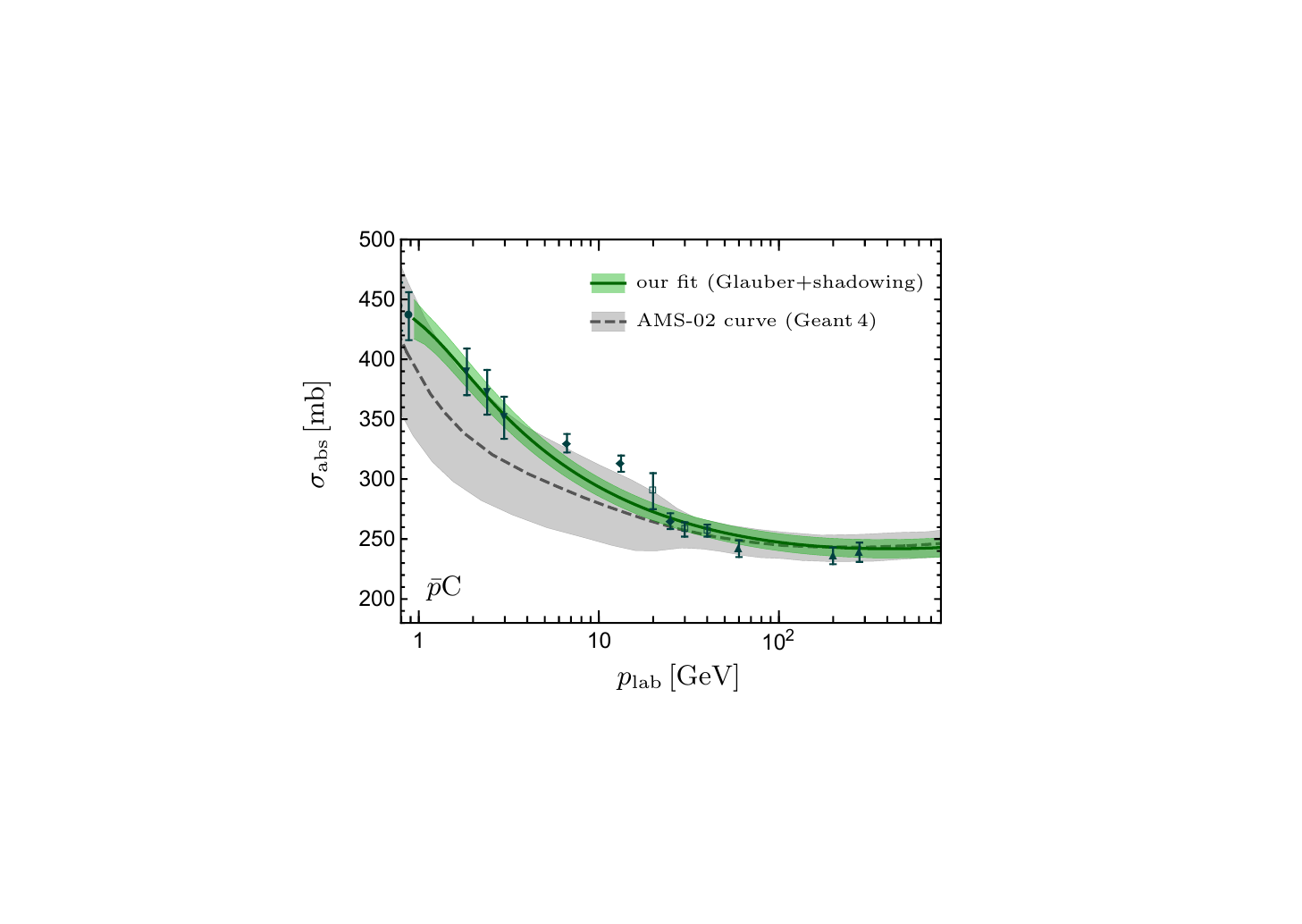}}
\end{picture}
\caption{
$\bar p$C  absorption cross section as a function of the projectile momentum $p_\text{lab}$. The solid dark green curve and green shaded band denote our best-fit cross section and $1\sigma$ uncertainty, respectively. Data points (containing $1\sigma$ error bars) of different experiments are denoted by individual symbols. For comparison, we show the cross section used in the AMS-02 analyses stemming from an implementation in Geant 4 (gray dashed curve and gray shaded band).
\label{fig:pbCpCxs}  
}
\end{figure*}

The most relevant process is absorption on carbon nuclei as the AMS-02 detector material mainly contains carbon ($\sim80\%$).
To compute the correlations in the cross-section uncertainties we perform a global fit to the $\bar p$C absorption cross section data~\cite{Allaby:1970pv,Abrams:1972ab,Denisov:1973zv,Carroll:1978hc,Nakamura:1984xw} within the above model varying all input parameters according to their nominal values and correlated uncertainties derived in a separate fit. The result is shown in Fig.~\ref{fig:pbCpCxs}.\footnote{We obtain analogous results for the $p$C absorption cross section entering the covariance matrices of the $\bar p/ p$ flux ratio and $p$ flux, see~\cite{Heisig:2020nse} for details.} We obtain the correlation matrix for $\sigma_\text{abs}$ (in terms of the $p_\text{lab}$-values that correspond to the rigidity bins of the AMS-02 data) directly from the statistical sample of the global fit.

%=====================
%    \                                           |
%      \                                         |
%        \                                       |
%                                      \         |
%                                        \       |
%                                          \     |
%=====================

%===================================================================
\section{Implications for dark matter}\label{sec:excess}
%===================================================================

To investigate the significance of a possible dark-matter signal in the AMS-02 antiproton data
we perform global cosmic-ray fits with and without a contribution from dark-matter annihilation. 
We model the latter by assuming annihilation into $\bar b b$ and considering the dark-matter mass and annihilation cross section as free parameters.
Following the approach of~\cite{Cuoco:2019kuu}, we use a minimal set of fluxes that allows us to simultaneously 
constrain the propagation and dark-matter model parameters, namely the $\bar p/p$ flux ratio~\cite{Aguilar:2016kjl} and the proton and helium fluxes~\cite{Aguilar:2015ooa,Aguilar:2015ctt}. 
We model the primary spectra by broken power laws and employ the standard diffusion coefficient with a slope $\delta$ but with additional freedom at low rigidities parametrized by $\eta$:
\begin{equation}\label{eq:eta}
D_{xx} \propto \beta^{\eta} \mathcal{R}^\delta\,.
\end{equation}
(Here, $\beta$ is the particle's velocity.) Negative values of $\eta$ resemble a low-rigidity break in diffusion that has recently been considered to improve the fit to secondary nuclear cosmic rays~\cite{Genolini:2019ewc,Weinrich:2020cmw}. 

We take into account convection and reacceleration and use GALPROP~\cite{Strong:2001gh} for the numerical solution of the diffusion equation. Solar modulation is implemented via the standard force-field approximation but with individual Fisk potentials for both charge signs. We exclude AMS-02 data for $\mathcal{R}<5\:\text{GV}$ in the fit where the approximation becomes questionable. To constrain solar modulation, we include Voyager data on protons and helium~\cite{Stone150}.
Uncertainties in the production cross sections of antiprotons are taken into account through the covariance matrix derived in~\cite{Reinert:2017aga}.

The correlations in the AMS-02 data are taken into account by summing up the individual covariance matrices for all contributions listed in Sec.~\ref{sec:systematic}. Of particular relevance are the correlations in the cross sections error discussed in Sec.~\ref{sec:absorption}. 
As the absorption-corrected fluxes scale inversely proportional to the cosmic-ray survival probability ($P = \E^{-n \sigma_\text{abs}}$, where $n$ accounts for the amount of detector material traversed) at linear order in the cross-section error, the correlation matrices for the measured fluxes
are identical to the ones for the cross sections.
Note that the correlation matrix for the flux ratio contains the contributions from both the $\bar p$C and $p$C cross section weighted by the relative magnitude of the individual uncertainties.

%=====================
%    \                                           |
%      \                                         |
%        \                                       |
\begin{figure*}[th]
\centering
\setlength{\unitlength}{1\textwidth}
\begin{picture}(1,0.433)
 \put(-0.0054,-0.02){\includegraphics[width=1\textwidth, trim= {1.55cm 13.9cm 1.7cm 5.7cm}, clip]{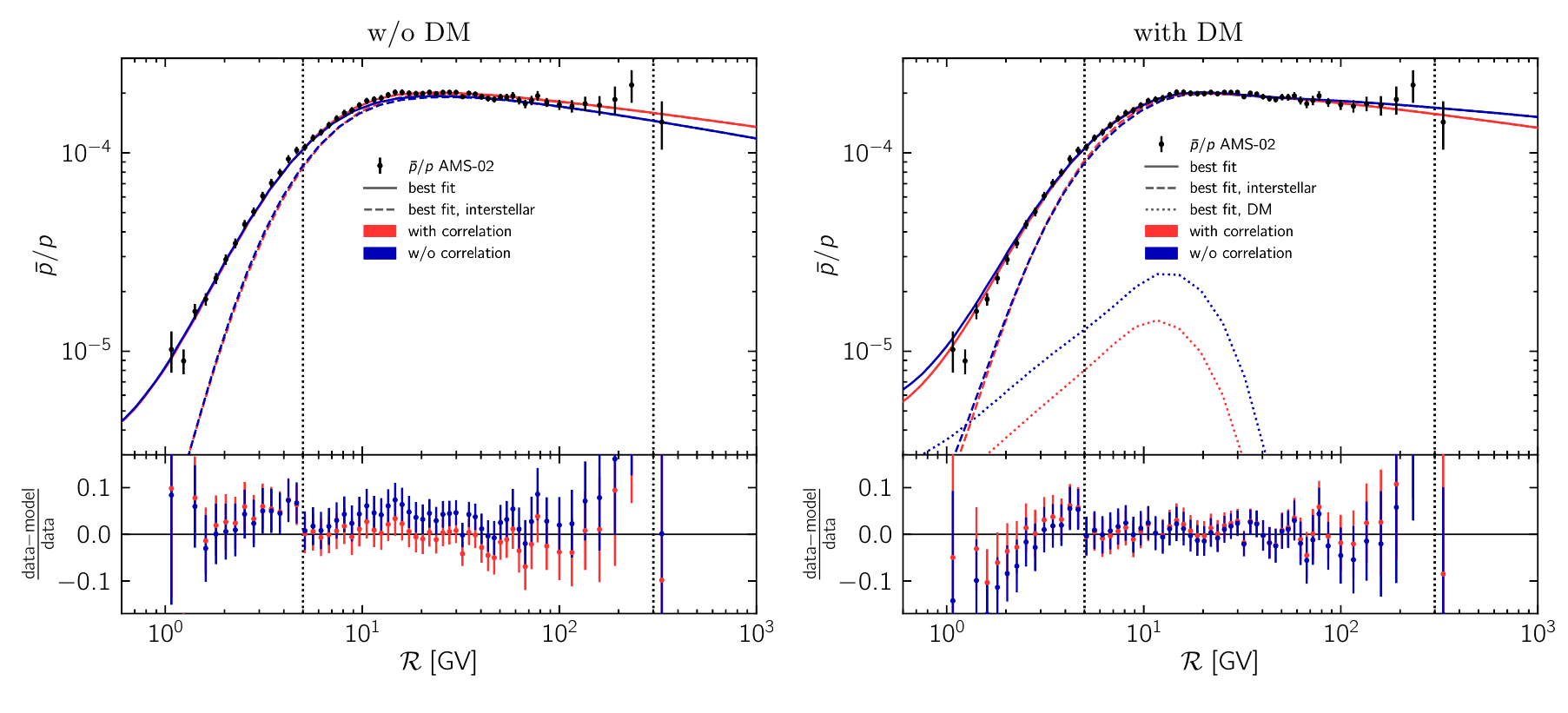}}
\end{picture}
\caption{ 
          Antiproton-to-proton flux ratio of the fit without (left) and with dark matter (right). 
          The solid red and blue curves denote the best-fit spectra at the top of the atmosphere with and without correlations in the AMS-02 errors, respectively. The dashed curves denote the corresponding interstellar fluxes. 
          The error bars of the AMS-02 data points denote the statistical and systematic uncertainties according to the diagonal entries of the total experimental covariance matrix.
       \label{fig:fit}
        }
\end{figure*}
%                                      \         |
%                                        \       |
%                                          \     |
%=====================

The best-fit results for the $\bar p/p$ ratio without (left) and with (right) dark matter are shown in Fig.~\ref{fig:fit}. To highlight their impact, we display the results without (blue) and with (red) the derived AMS-02 correlation matrices. The corresponding global significance of the excess drops from $1.8\,\sigma$ to $0.5\,\sigma$. Accordingly, we do not find any significant preference for dark matter in the data.

However, the situation is not fully conclusive. Considering the involved (correlated) uncertainties and the effect of the diffusion parameter $\eta$ one by one, we can make an interesting observation. Taking into account the AMS-02 correlation matrices only (no antiproton production cross section uncertainties and setting $\eta=1$) increases the significance to above $5\sigma$. This is because errors are no longer treated uncorrelated and hence the fit provides less freedom to accommodate spectral features that are not in line with those potentially caused by the correlations. Taking into account the uncertainties from antiproton production cross section in addition yields a significance slightly above $3\sigma$ while only the inclusion of these uncertainties plus the extra freedom in the diffusion coefficient at low rigidities allows us to reconcile the tension in the data without introducing dark matter. Hence, the presence of the excess decisively depends on the diffusion model at low rigidities. Interestingly, uncertainties in nuclear cross sections are, again, a limit factor in establishing the diffusion model~\cite{Korsmeier:2021brc}.

Finally, we note that a more recent measurement of the 7-year $\bar p$ flux by AMS-02 \cite{AMS:2021nhj} shows systematic deviations from the previous measurement at around 20 GV which potentially decrease the significance of the excess even further \cite{DiMauro:2021qcf}.

%===================================================================
\section{Conclusions}\label{sec:concl}
%===================================================================

The AMS-02 antiproton data provides us with a powerful tool to search for dark-matter annihilation in our Galaxy.
Interestingly, data exhibits a residual component at a rigidity $\mathcal{R}\sim 10\!-\!20\:\text{GV}$ that 
has been interpreted as a signal from dark-matter annihilation with a mass around 100\,GeV\@.
However, systematic uncertainties at few percent level become important
such as the antiproton production cross section uncertainties. 

Here, we have considered another source of uncertainties, namely
correlations in the AMS-02 systematic errors that potentially have a large effect on the significance of the signal. 
The most relevant of these stem from cross sections for cosmic-ray absorption in the detector for which we computed the full covariance for the first time.
These correlations are vital to fully exploit the precision of the experiment. Their inclusion reveals that the 
excess is not robust. However, the significance of the signal decisively depends on the diffusion model  at low rigidities -- a subject currently under active investigations.
Again, uncertainties in nuclear cross sections that enter its inference are the limiting factor and motivate dedicated experimental efforts in the future.

\providecommand{\href}[2]{#2}\begingroup\raggedright\endgroup

\end{document}